\RequirePackage{fix-cm}

\documentclass[smallextended]{svjour3}

\smartqed  

\usepackage{cite}
\usepackage{booktabs}
\usepackage{enumerate}
\usepackage{graphicx}
\graphicspath{{figures/}}
\usepackage{multirow}
\usepackage{framed}
\usepackage{lipsum}
\usepackage{color}
\usepackage{amsmath}
\usepackage{courier}
\usepackage{hyperref}
\usepackage{natbib}
\usepackage[framemethod=TikZ]{mdframed}
\usepackage{mathptmx}

\usepackage{caption} 
\newcommand{\rqone}{What are the aggregation techniques that do not inflate the correlation between LOC and other metrics at the file level?}
\newcommand{\rqtwo}{Do different aggregation techniques convey different information?}
\newcommand{\rqthree}{Does the type of aggregation used in defect prediction models matter?}

\begin{document}

\title{Investigating the Impact of Metric Aggregation Techniques on Defect Prediction}

\author{Rawad Abou Assi}

\institute{School of Computing, Queen's University, Canada \\
              \email{rawad@cs.queensu.ca}
}

\date{ }

\maketitle

\begin{abstract}
Code metrics collected at the method level are often aggregated using summation to capture system properties at higher levels (e.g., file- or package-level).
Since defect data is often available at these higher levels, this aggregation allows researchers to build defect prediction models.
Recent findings by Landman et al. indicate that aggregation is likely to inflate the correlation between size and complexity metrics.
In this paper, we explore the effect of nine aggregation techniques on the correlation between three types of code metrics, namely Lines of Code, McCabe, and Halstead metrics.
In addition to summation, we study aggregation techniques that are measures of: (1) central tendency (average and median), (2) dispersion (standard deviation and inter-quartile range), (3) shape (skewness and kurtosis), and (4) income inequality (Theil index and Gini coefficient). Our results show that defect prediction models built using summation outperform those built using other aggregation techniques. We also find that more complex aggregations are no different than much simpler ones and that incorporating all aggregation types in the same model does not provide a significant improvement over using summation alone.
\end{abstract}

\section{Introduction}
Code metrics can be computed at different levels of granularity to help analyze the defect-proneness of software components.
While metrics such as the Chidamber and Kemerer suite~\citep{chidamber1994} are defined at the class-level, others such as Lines Of Code (LOC), \citep{mccabe1976}, and \citep{halstead1977} measure complexity at the module level.
These module-level metrics are typically aggregated using sum or average to lift their values to higher levels~\citep{gill1991, lanza2006, manet2011}.

Recent findings of \cite{landman2014} suggest that aggregation is likely to amplify the correlation between LOC and other code metrics at the file level, which is consistent with previous results \citep{li1987, jay2009, basili1984, feuer1979, curtis1979b}.
Their study, which involved a large corpus of Java methods, indicates that the linear correlation between LOC and McCabe's cyclomatic complexity increases when both metrics are summed over larger units of code.
However, it is not known whether this observation generalizes to other types of metrics or other aggregation techniques.

We, therefore, aim to study the impact that different aggregation techniques have on defect prediction at the file level using three types of code metrics, namely LOC, McCabe, and Halstead metrics.
In addition to summation, we study aggregation techniques that measure:
(1) central tendency (\textit{Average} and \textit{Median}),
(2) dispersion (\textit{Standard Deviation} and \textit{Inter-quartile Range}),
(3) shape (\textit{Skewness} and \textit{Kurtosis}),
and (4) income inequality (\textit{Theil} index and \textit{Gini} coefficient).
We analyze the aggregated metrics by performing correlation analysis and building regression models with repeated cross-validation.
Our empirical study involves 12 releases of three open-source projects and addresses the following research questions:

\begin{enumerate}[{\bf (RQ1)}]

\item {\bf \rqone}\\
  As reported by \cite{landman2014}, we find that aggregating the studied complexity metrics using \textit{Sum} tends to inflate their correlation with LOC at the file level.
 On the other hand, the correlation with LOC tends to decrease when aggregation is performed using \textit{Median} as well as the measures of shape and income inequality.

\item {\bf \rqtwo}\\
  Yes, although we observe high rates of correlation among aggregations of the same family (e.g., \textit{Skewness} and \textit{Kurtosis}, \textit{Theil} and \textit{Gini}), the redundancy among aggregations of different families remains low.

\item {\bf \rqthree}\\
Using repeated 10-fold cross validation, our results show that defect prediction models built using summation outperform those built using other aggregation techniques. We also conclude that incorporating all aggregation types in the same model does not provide a significant improvement over using summation alone.

\end{enumerate}

The remainder of the paper is organized as follows.
Section~\ref{sec:related} discusses related work within the context of metric aggregation.
Section~\ref{sec:setup} describes our experimental setup by elaborating on the code metrics and aggregation techniques that we use.
It also presents our regression models along with the evaluation criteria.
Section~\ref{sec:results} presents a detailed analysis of the results.
Section~\ref{sec:threats} discusses the threats to the validity of our findings.
Finally, Section~\ref{sec:conclusions} concludes our work and points out possible future enhancements. 

\section{Related Work}
\label{sec:related}
There seems to be a trade-off between the simplicity of an aggregation method and its suitability to represent the data accurately. While the former is important to arrive at an easy-to-describe aggregation, the latter is necessary to provide reliable decisions about software maintainability. For example, even though \textit{Sum} is a simple metric, it was recently shown to render some code metrics redundant \citep{landman2014}. Also, simple measures of central tendency such as the mean are often deemed inappropriate when the underlying data is skewed \citep{concas2007, vasa2009}. As a result, several aggregation approaches were proposed in literature to overcome such limitations.

One family of metric aggregation techniques uses income inequality measures that were originally proposed to quantify the imbalance in wealth distribution in a given population. In this regard, \cite{vasa2009} use the Gini coefficient \citep{gini1912} on class-level metrics to understand the evolution of object-oriented systems. Their results indicate that the Gini coefficients do not change significantly between adjacent releases and that the relatively high values of such coefficients indicate that developers tend to prefer centralized and complex abstractions over distributed and simple ones. On the other hand, \cite{serebrenik2010} propose using Theil index \citep{theil1967} instead of the Gini coefficient for metric aggregation as the latter is not decomposable among separate groups of a given population. They argue that decomposability highlights intra-group and inter-group inequality and thus is essential to explain the inequality among a given population rather than just measuring it.

Another approach for metric aggregation relies on fitting a statistical distribution to the set of observed metric values. In this sense, the aggregation would be based on the estimated values of the distribution parameters. It is evident that the viability of such approach is dependent on the choice of statistical distribution as well as the dataset being used. For example, while \cite{tamai2002} report that size data such as the number of methods per class and the number of lines of code per method follow a negative binomial distribution, \cite{concas2007} argue that the former follows a log-normal distribution and the latter follows a power-law.

\section{Experimental Design}
\label{sec:setup}
Our study involves 12 releases of three open source projects: \textit{Eclipse}, \textit{Apache Ant}, and \textit{jEdit}. \textit{Eclipse} (eclipse.org) is a cross-platform IDE with an extensible plug-in architecture. \textit{Ant} (ant.apache.org) is a build tool implemented in Java - initially proposed as an alternative to Unix \textit{make}. \textit{jEdit} (jedit.org) is a Java-based text editor targeted for programmers. Table \ref{projects} shows, for each project, the releases that we used as well as the download locations where we obtained the corresponding source code and post-release defect information.

\begin{table}[h]
\centering
\caption{Studied Projects.}
\begin{tabular}{llll}
\hline
 & Releases    & Source code & Post-release defects \\
\hline
Eclipse & 2.0, 2.1, 3.0 & archive.eclipse.org & st.cs.uni-saarland.de\\
Ant & 1.3, 1.4, 1.5, 1.6, 1.7 & archive.apache.org & code.google.com/p/promisedata\\
jEdit & 4.0, 4.1, 4.2, 4.3 & sourceforge.net & code.google.com/p/promisedata\\
\hline
\end{tabular}
\label{projects}
\end{table}

After obtaining the source code of the 12 releases, we used McCabe IQ 8.3 \citep{mccabeiqwebpage} to collect three types of method-level code metrics: LOC, McCabe, and Halstead  metrics. These metrics are then aggregated to the file level via nine aggregation techniques. Using the aggregated metrics and the post-release defect information, we build regression models to study the impact of the different aggregations on defect prediction. Figure 1 shows an overview of our approach.

\begin{figure}[h]
  \centering
    \includegraphics[width=1\textwidth]{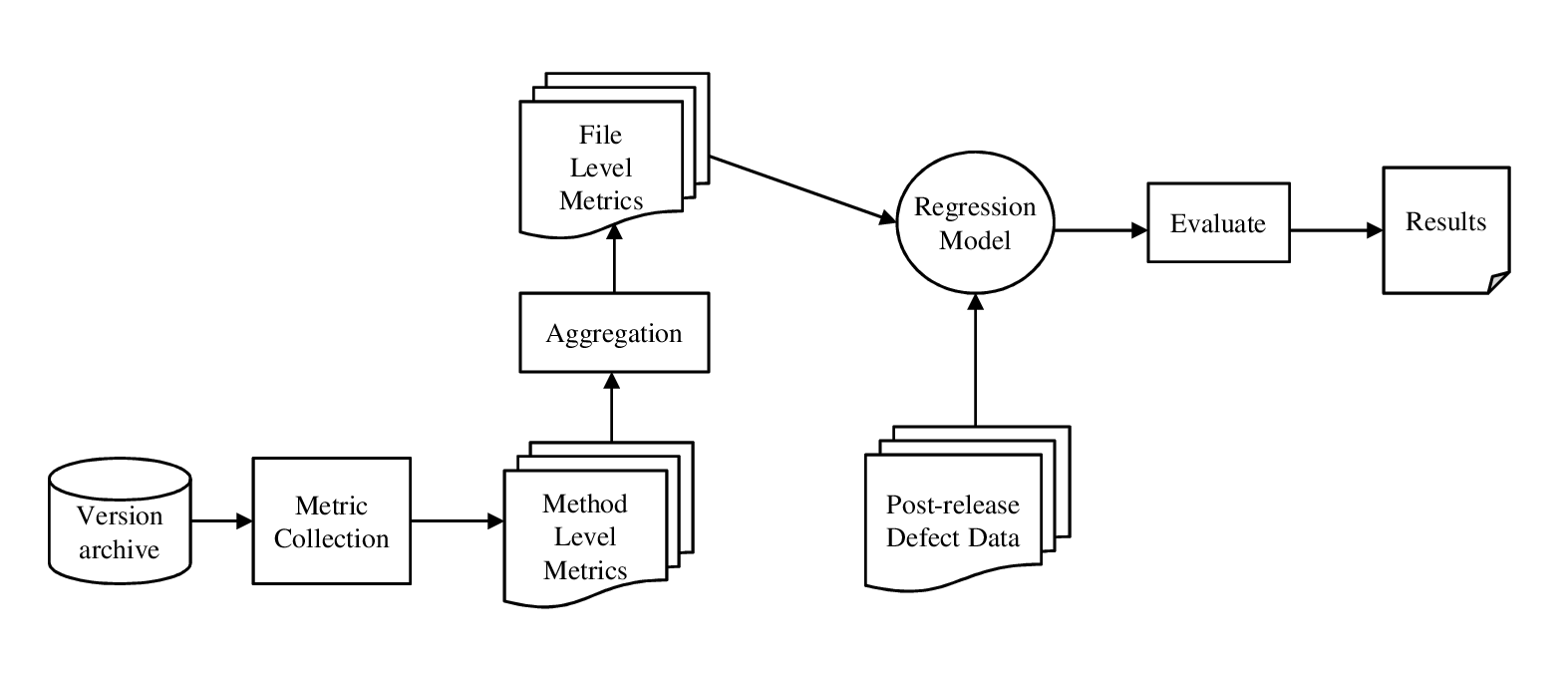}
    \caption{Approach overview.}
\end{figure}
  
\subsection{Code Metrics}
Code metrics are relatively easy to compute and are widely used in literature \citep{sunohara1981, menzies2004, zhang2009, zimmermann2007, jiang2008, curtis1979a, menzies2007, khoshgoftaar1990} to assess software quality, build prediction models, understand maintenance complexity, etc. In this work, we consider three types of commonly used code metrics: LOC, McCabe, and Halstead metrics. While McCabe metrics such as cyclomatic complexity \citep{mccabe1976} aim at measuring the complexity associated with a software module using the number of forks in the corresponding control-flow graph, Halstead metrics estimate the complexity using the number of operators and operands in the module.
Table \ref{metrics} lists the metrics we use in our study and provides a brief description of each as described in \cite{mccabeiqwebpage}.

\begin{table}[t]
\centering
\caption{Code metrics used.}
\begin{tabular}{p{0.4\textwidth}p{0.1\textwidth}p{0.4\textwidth}}
\hline
Metric    & Type & Description\\
\hline\\[2pt]
LOC      &  -   & Total number of lines in a module\\[7pt]
Cyclomatic Complexity (v(G))     &  McCabe   & Number of linearly independent paths in the flow-graph of a module\\[7pt]
Essential Complexity (ev(G))      &  McCabe   & Degree to which a module contains unstructured constructs\\[7pt]
Module Design Complexity (iv(G))     &  McCabe   & Complexity of the design reduced module\\[7pt]
Program Volume (V)      &  Halstead & Minimum number of bits required for coding the program\\[7pt]
Program Difficulty (D)      &  Halstead &  Level of difficulty in the program\\[7pt]
Program Length (N)      &  Halstead   & Total number of operators and operands\\[7pt]
Program Level (L)      &  Halstead & Level at which the program can be understood\\[7pt]
Intelligent Content (I)      &  Halstead & Complexity of a given algorithm independent of the language used\\[7pt]
Programming Effort (E)      &  Halstead & Estimated mental effort required to develop the program\\[7pt]
Error Estimate (B)      &  Halstead &  Estimated number of errors in the program\\[7pt]
Programming time (T)      &  Halstead   & Estimated amount of time to implement the algorithm\\[7pt]
\hline
\end{tabular}
\label{metrics}
\end{table}

\subsection{Aggregation Techniques}
For a given module $D$ and a code metric $M$, we denote by $M(D)$ the value of $M$ when applied to $D$. For example $LOC(foo)$ is the number of lines of code in method $foo$. If a file $F$ contains $N$ methods ${D_{1}, D_{2},...,D_{N}}$, then $M$ could be defined for $F$ by aggregating the values $M(D_{1}), M(D_{2}),...,M(D_{N})$. Such aggregation could be done in different ways, giving rise to different interpretations of $M$ at the file level. For example, aggregating $LOC$ by \textit{Sum} would result in a file-level metric (say \textit{Sum(LOC)}) that measures the overall size of $F$. However, aggregating it using \textit{Standard Deviation} would result in a different file-level metric (\textit{SD(LOC)}). In this work, we consider nine aggregation techniques including the typical \textit{Sum}. For a given set of values $X=\{x_{1}, x_{2}, ..., x_{N}\}$ we define the nine aggregation techniques as follows. Note that $Q_{1}(X)$, $Q_{2}(X)$, and $Q_{3}(X)$ denote the first, second, and third quartiles of X respectively.

\subsubsection{Summation}
$Sum(X)=\sum\limits_{i=1}^N x_{i}$

\subsubsection{Measures of Central Tendency}

\begin{itemize}
\item[--] Average: $Avg(X)=\bar{X}=\frac{Sum(X)}{N}$
\item[--] Median: $Med(X)=Q_{2}(X)$
\end{itemize}

\subsubsection{Measures of Dispersion}

\begin{itemize}
\item[--] Standard deviation: $SD(X)=\sqrt{\frac{\sum\limits_{i=1}^N (x_{i}-\bar{X})^{2}}{N-1}}$\\
\item[--] Inter-quartile range: $IQR(X)=Q_{3}(X)-Q_{1}(X)$
\end{itemize}

\subsubsection{Measures of Shape}
\begin{itemize}
\item[--] Skewness: $Skew(X)=\frac{\frac{\sum\limits_{i=1}^N (x_{i}-\bar{X})^3}{N}}{(\frac{\sum\limits_{i=1}^N (x_{i}-\bar{X})^2}{N-1})^{\frac{3}{2}}}$\\
\item[--] Kurtosis: $Kurt(X)=\frac{\frac{\sum\limits_{i=1}^N (x_{i}-\bar{X})^4}{N}}{(SD(X))^{4}}-3$
\end{itemize}

\subsubsection{Income Inequality Measures}
\begin{itemize}
\item[--] Theil index: $Theil(X)=\frac{1}{N}\sum\limits_{i=1}^N (\frac{x_{i}}{\bar{X}}ln\frac{x_{i}}{\bar{X}})$\\
\item[--] Gini coefficient: $Gini(X)=\frac{2\sum\limits_{i=1}^N ix'_{i}}{N\sum\limits_{i=1}^N x'_{i}}-\frac{N+1}{N}$\\
where $\{x'_{1}, x'_{2}, ..., x'_{N}\}$ is a sorted version of $X$.
\end{itemize}

\subsection{Selection of Predictors}
Given that we have 12 metrics and 9 aggregation techniques, it follows that there are 108 predictors to account for in our regression models. However, prior to building the models, it is important to mitigate any potential redundancy among these predictors to avoid model instability and degradation in predictive performance \citep{kuhn2013}. Redundancy among predictors exists in two forms: pairwise correlation and multi-collinearity. While correlation measures the degree of association between pairs of variables, multi-collinearity refers to the situation when there is a concurrent relationship between multiple variables, i.e. when some variables could be predicted from a combination of other ones. In this work, we pre-process the set of predictor variables by applying correlation and multi-collinearity analyses as follows: 
\begin{enumerate}
  \item Correlation analysis: We use \textit{Varclus} \citep{sarle1990} to analyze correlations among the predictor variables. \textit{Varclus} is a hierarchical approach that depicts variables in clusters each of which is associated with a correlation level. We consider Spearman rank correlation ($\rho$) such that, for each cluster with $|\rho|>0.7$, we select one variable and discard all the rest. We use the \textit{varclus} function provided by the \textit{Hmisc} package in R \citep{rwebpage} to perform this analysis.
  \item Multi-collinearity analysis: The variables retained from the previous step are further analyzed to check for potential multi-collinearity. We perform such analysis by considering how well each variable is predicted by the others. Variables that are highly predictable are discarded in an iterative fashion until all remaining variables exhibit low levels of redundancy. Specifically, we quantify the predictability of a particular variable \textit{v} by fitting a linear regression model that uses \textit{v} as response and the other variables as predictors. We then use the adjusted coefficient of determination $R^2$, which represents the goodness of fit, to measure the predictability of \textit{v}. Therefore, in each iteration, we compute the $R^2$ for all variables and drop the one associated with the highest value. This process is repeated until all the $R^2$ values are below a cutoff threshold. This approach is implemented via function \textit{redun} provided by the \textit{Hmisc} package in R, which uses a default cutoff threshold of 0.9.  
\end{enumerate}
We refer to the pre-processing stage comprising these two steps as the "filtering" phase since it is used to filter the predictor variables that are used in the models. Since this  phase could be applied on the method-level metrics as well as the aggregated ones, we propose two alternatives for selecting the predictor variables, which we describe next.

\subsubsection{One-level filtering approach}
\begin{figure}[h]
  \centering
    \includegraphics[width=1\textwidth]{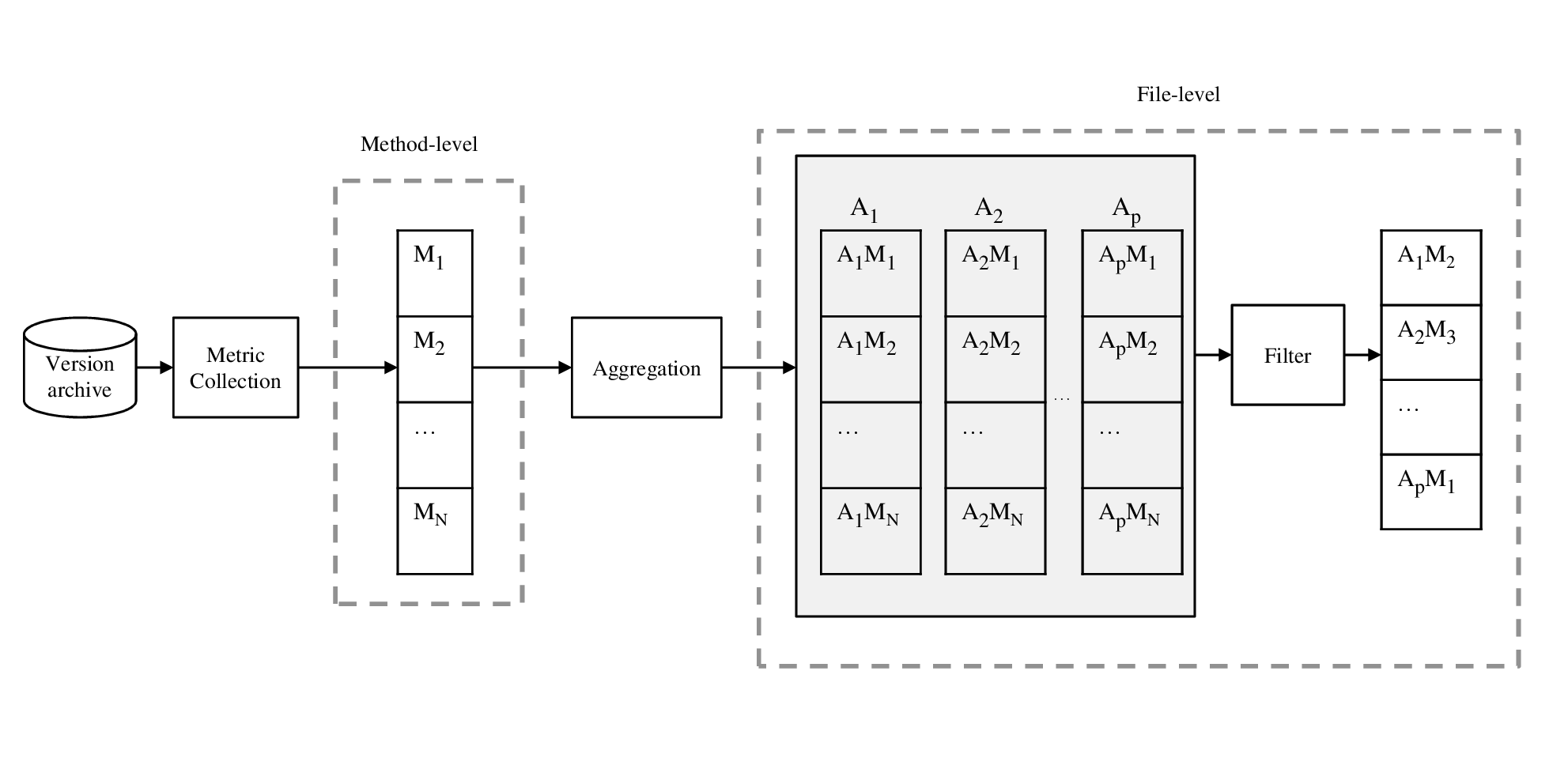}
    \caption{One-level filtering approach (the final set of variables is depicted randomly).}
    \label{Filter1}
\end{figure}

In this approach, we first perform aggregation to obtain the file-level metrics. Then, we apply the filtering mechanism on the resulting 108 variables. Figure \ref{Filter1} shows an overview of this approach where the method-level metrics are denoted by $M_1$, $M_2$, ..., $M_N$ and the aggregation functions by $A_1$, $A_2$, ..., $A_P$.

\subsubsection{Two-level filtering approach}
\begin{figure}[h]
  \centering
    \includegraphics[width=1\textwidth]{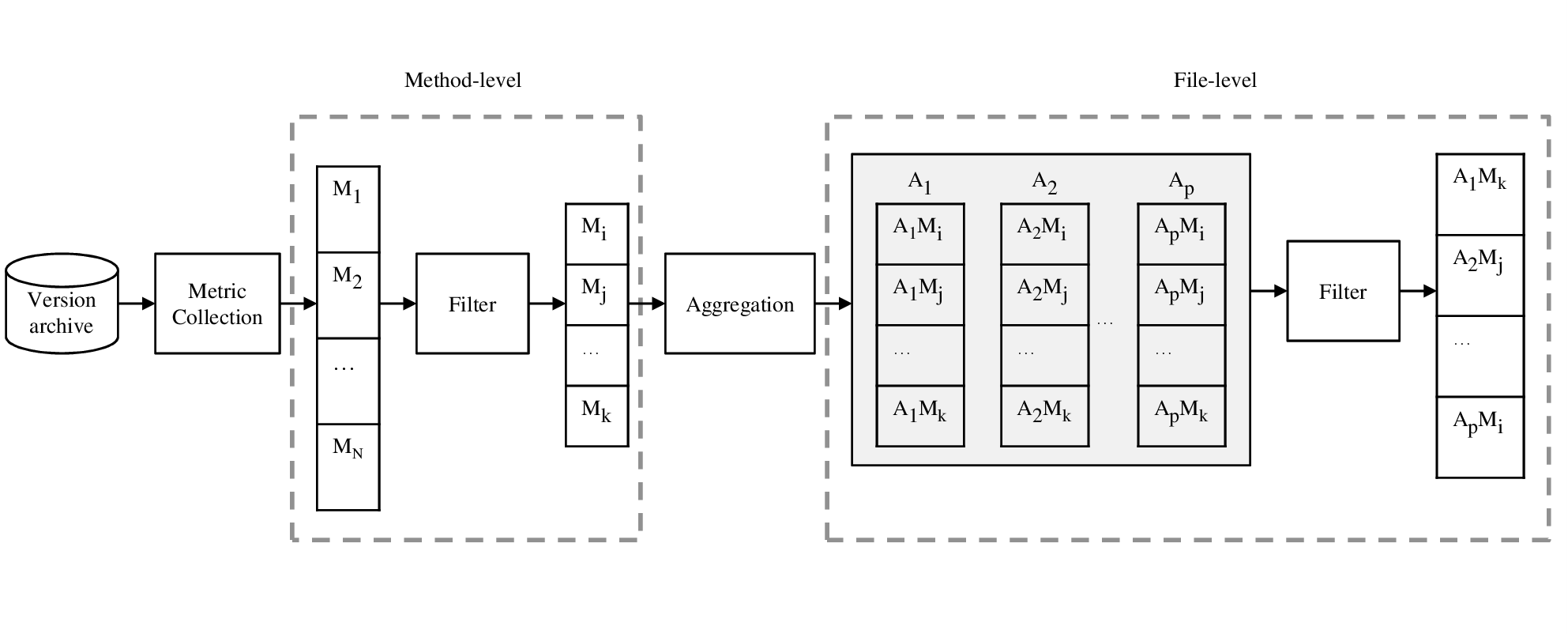}
    \caption{Two-level filtering approach (the final set of variables is depicted randomly).}
\label{Filter2}
\end{figure}

This approach is depicted in Figure \ref{Filter2}. In contrast to the previous approach, we first apply filtering on the method-level metrics to obtain $M_i$, $M_j$, ..., $M_k$. These metrics are then aggregated to the file level using each of the aggregation techniques. At the file level, we re-apply filtering to obtain the final set of predictor variables.

\subsection{Defect Prediction using Aggregated Code Metrics}
Defect prediction aims at modeling the relationship between a set of explanatory (i.e. predictor) variables and the fault-proneness of a software component (dependent variable). In this work, we use as predictor variables the set of aggregated metrics resulting from the two aforementioned filtering approaches. We study the impact of such metrics on the accuracy of defect prediction by building linear and logistic regression models having the file-level defect information as the dependent variable.   

\subsubsection{Regression Models}
Linear regression uses a linear equation of the form $\beta_0 + \beta_1x_1 + ... + \beta_nx_n$ to predict the value of the dependent variable, where $x_1 , ... , x_n$ are assumed to be the predictor variables. The regression coefficients $\beta_0, \beta_1, ..., \beta_n$ are usually determined relative to a given training set using \textit{Least-squares} estimation. In our case, we use linear regression to predict the number of post-release defects that are likely to be present in a given file.

As opposed to linear regression, the dependent variable in logistic regression is a binary variable, indicating whether a file has defects or not. This outcome is predicted using a logistic function of the form $\frac{1}{1+e^{-(\beta_0 + \beta_1x_1 + ... + \beta_nx_n)}}$. The output would be a value between 0 and 1. Typically, a file would be classified as defective if the output is greater than $0.5$.

In our analysis, we use the implementations \textit{lm} and \textit{glm} provided by R \citep{rwebpage} to build the linear and logistic models respectively.

\subsubsection{Performance Measures}
We evaluate the predictive accuracy of our regression models using repeated 10-fold cross validation \citep{witten2005}. Specifically, we partition each dataset $D$ into 10 equally-sized disjoint subsets $S_1, S_2, ..., S_{10}$. For each subset $S_i$, we use $D-S_i$ to build the model and $S_i$ for prediction. The whole process is then repeated 10 times to account for the effect of randomness.

In each iteration, we measure the accuracy of the linear models by computing the mean squared error (MSE) between the predicted values and the actual ones. As for the logistic models, we use the area under the ROC curve \citep{bradley1997} (AUC) which is commonly used to evaluate binary classifiers. AUC ranges between 0 and 1 where higher values indicate better performance. As such, a random classifier would have an AUC of 0.5. We report the average values for MSE and AUC metrics across all the iterations of the cross-validation.

\section{Results}
\label{sec:results}
In this section, we present the results of our empirical study with respect to our three research questions.\\
\\
\textbf{(RQ1) \rqone}\\
\\In an ideal situation, one would desire aggregation techniques that preserve as much information as possible about the signals exhibited by method-level metrics when lifting their values to the file level. However, previous research indicates that aggregating certain metrics to the file level using summation inflates their correlation with LOC \citep{landman2014, li1987, jay2009, basili1984, feuer1979, curtis1979b}. This might have a detrimental effect as many metrics might become redundant at the file level. This research question aims at investigating how aggregation techniques other than the typical summation affect the correlation with LOC at the file level. We address this issue by comparing correlations before and after aggregation. Specifically, we first compute the Spearman rank correlation $\rho$ between $LOC$ and every other metric at the method level. Then, for each aggregation technique $A$ and each metric $M$, we compute $\rho$ for $A(LOC)$ and $A(M)$. In our analysis, we consider the absolute value of $\rho$ since we are interested in the degree of correlation rather than its polarity. Table \ref{CorrelationIncrease} reports the average increase in $|\rho|$ across all 12 datasets between $LOC$ and each code metric, according to the nine aggregation functions. For example, the average increase in $|\rho|$ for LOC and cyclomatic complexity when \textit{Sum} is used for aggregation is 14\%.  

\begin{table}[htbp]
  \centering
  \caption{Average increase in Spearman rank correlation with LOC at the file level.}
    \begin{tabular}{rrrrrrrrrr}
    \toprule
          & Sum   & Avg   & Med   & SD    & IQR   & Skew  & Kurt  & Theil & Gini \\
    \midrule
    v(G)  & 14\%  & 4\%   & -5\%  & 4\%   & 3\%   & 1\%   & 0\%   & 0\%   & 2\% \\
    ev(G) & 83\%  & 27\%  & -28\% & 38\%  & 4\%   & 27\%  & 35\%  & 26\%  & 26\% \\
    iv(G) & 16\%  & 5\%   & -7\%  & 6\%   & 2\%   & 1\%   & -1\%  & 1\%   & 3\% \\
    N     & 6\%   & 4\%   & 0\%   & 4\%   & 3\%   & 0\%   & -1\%  & -5\%  & -5\% \\
    V     & 5\%   & 4\%   & 0\%   & 4\%   & 3\%   & 0\%   & -1\%  & -11\% & -12\% \\
    L     & -10\% & -24\% & -5\%  & -77\% & -87\% & -51\% & -8\%  & -31\% & -24\% \\
    D     & 8\%   & -1\%  & -3\%  & -5\%  & -5\%  & -7\%  & -3\%  & -25\% & -21\% \\
    I     & 9\%   & -3\%  & -1\%  & -4\%  & -4\%  & -8\%  & -2\%  & -22\% & -18\% \\
    E     & 1\%   & 1\%   & 0\%   & 2\%   & 2\%   & -1\%  & -2\%  & -18\% & -19\% \\
    B     & 5\%   & 4\%   & 0\%   & 4\%   & 3\%   & -1\%  & -1\%  & -23\% & -20\% \\
    T     & 1\%   & 1\%   & 0\%   & 2\%   & 2\%   & -1\%  & -2\%  & -18\% & -19\% \\
    \bottomrule
    \end{tabular}%
  \label{CorrelationIncrease}%
\end{table}%

The results clearly show that the increase in correlation depends on the metric being considered as well as the aggregation function used. For all metrics, except Halstead's program level (L), there is at least one aggregation function that increases its correlation with LOC and another one that decreases it. Moreover, it can be noticed that most aggregation functions other than \textit{Sum} generally tend to decrease the correlation rather than increase it, especially for the Halstead metrics.

Our conclusions regarding \textit{Sum} are in line with previous research in the sense that it tends to increase correlation at the file level. However, we find this increase rather moderate as it ranges between $1\%$ and $16\%$ for all metrics except McCabe's essential complexity (ev(G)) and Halstead's program level (L).  Also, for most of the metrics, \textit{Sum} results in the highest increase among all the aggregation functions.

Concerning \textit{Avg}, \textit{SD}, and \textit{IQR}, we notice a similar pattern to that associated with \textit{Sum} although they generally result in lower increase rates. In contrast to \textit{Sum}, all of them decrease the correlation for Halstead's program difficulty (D) and intelligent content (I).

On the other hand, the results indicate that the overall tendency among \textit{Med}, measures of shape (\textit{Skew} and \textit{Kurt}), and income inequality measures (\textit{Theil} and \textit{Gini}) is to decrease the correlation with LOC at the file level rather than increasing it. One might argue that this would be as misleading as increasing the correlation. However, we believe it is not harmful as long as it does not render file-level metrics redundant while their method-level counterparts are not. In this regard, we notice that \textit{Med} is the only aggregation that does not inflate the correlation for any metric and, even for the cases where it decreases the correlation, we find that the decrease is generally low. With the exception of essential complexity ev(G), we can draw similar conclusions regarding the measures of shape. Concerning the income inequality measures, the overall decrease associated with them is more significant as it ranges between $5\%$ and $31\%$ for \textit{Theil} and between $5\%$ and $24\%$ for \textit{Gini}.

\begin{mdframed}[roundcorner=10pt]
\textit{Sum increases file-level correlation between most of the studied metrics and LOC. Conversely, \textit{Median} as well as the measures of shape and income inequality tend to decrease such correlation.}
\end{mdframed}
\vspace{5mm}

\begin{figure}[h]
  \centering
    \includegraphics[width=1\textwidth]{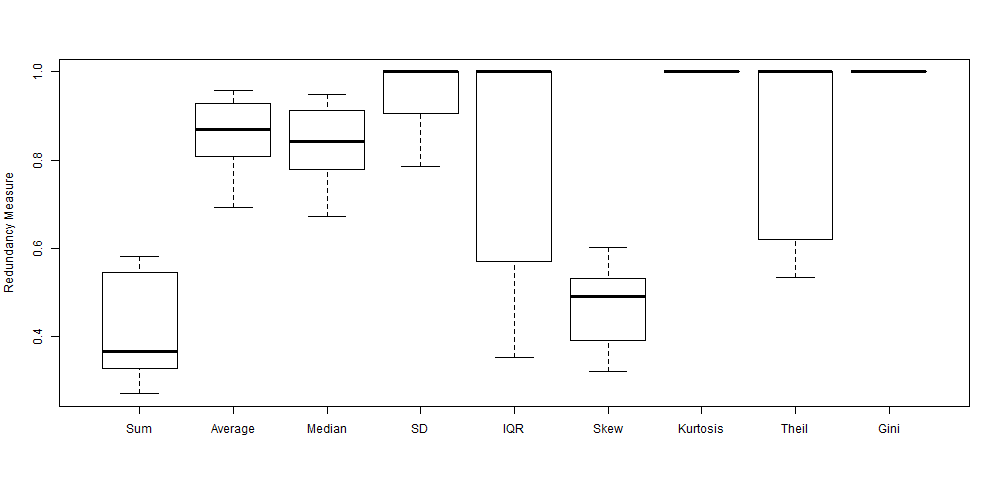}
    \caption{The distribution of the redundancy measures for each aggregation of LOC across the 12 datasets.}
\label{boxplot_LOC}
\end{figure}

\begin{figure}[h]
  \centering
    \includegraphics[width=1\textwidth]{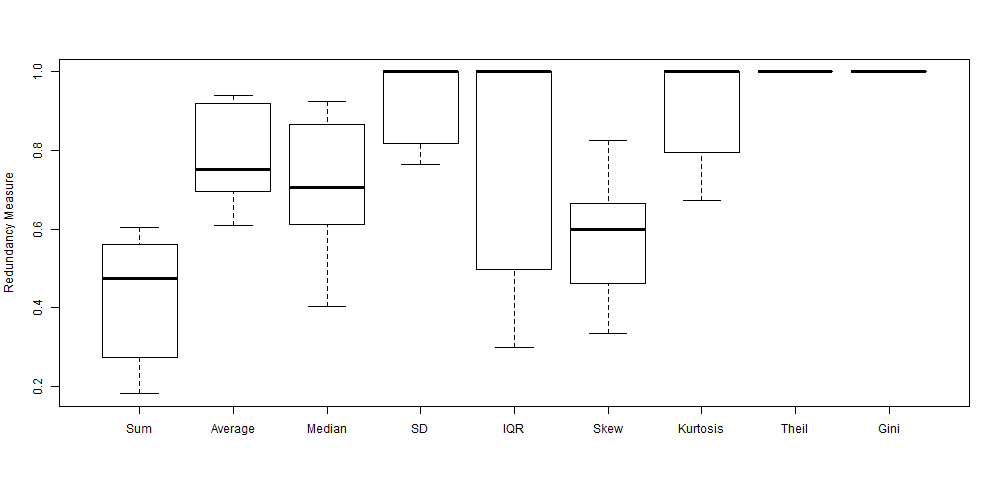}
    \caption{The distribution of the redundancy measures for each aggregation of McCabe's cyclomatic complexity v(G) across the 12 datasets.}
\label{boxplot_VG}
\end{figure}

\begin{figure}[h]
  \centering
    \includegraphics[width=1\textwidth]{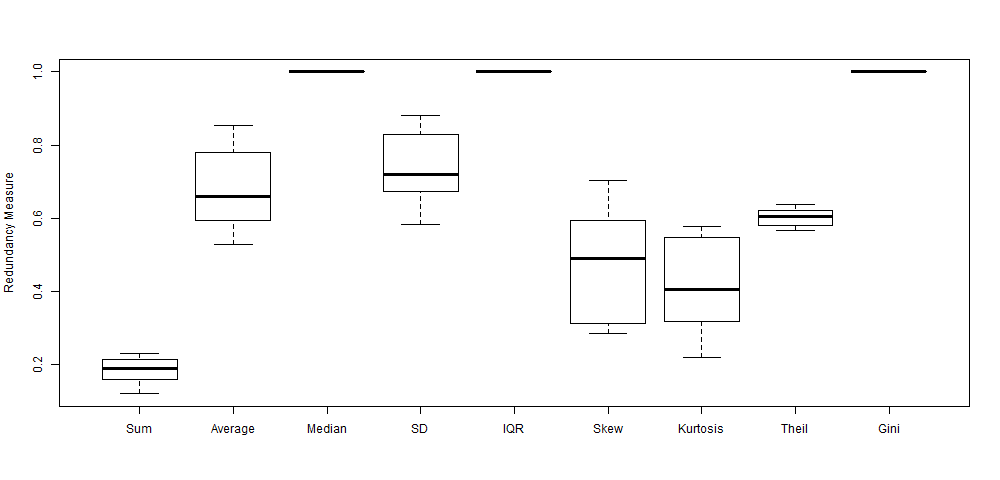}
    \caption{The distribution of the redundancy measures for each aggregation of Halstead's program level (L) across the 12 datasets.}
\label{boxplot_L}
\end{figure}

\textbf{(RQ2) \rqtwo}\\
\\In the previous research question, we investigated the redundancy of different file-level aggregated code metrics with respect to LOC. Nevertheless, it is also important to study different forms of redundancy among the aggregation techniques themselves; i.e., to check whether aggregating the same metric using different techniques would yield different signals. If this is the case, then one would expect to build defect prediction models with better explanatory power by incorporating all such aggregations if possible. We address this research question by assigning to each aggregated metric a "redundancy measure" that ranges between 0 and 1 where higher values indicate higher redundancy. We compute such redundancy measure using correlation and multi-collinearity analysis in a similar way to that presented in Section 3.3. For each code metric $M$, we quantify the redundancy associated with each aggregation function $A_i$ as follows:
\begin{enumerate}
  \item Correlation analysis: We apply \textit{Varclus} on the set ${A_1(M), A_2(M), ..., A_9(M)}$ and choose only one metric from each cluster having $|\rho|>0.7$. The metric to be chosen is the one that has the lowest rank in the following order: \textit{Sum}, \textit{Avg}, \textit{Med}, \textit{SD}, \textit{IQR}, \textit{Skew}, \textit{Kurt}, \textit{Theil}, and \textit{Gini}. For example, if the cluster contains \textit{Theil(M)}, \textit{Med(M)}, and \textit{IQR(M)}, we choose \textit{Med(M)} to be retained and discard all the rest. The order we devise is based on our judgment regarding the complexity associated with each aggregation. For example, we believe that measures of central tendency are simpler than measures of dispersion which are simpler than the measures of shape. We also consider \textit{Sum} to be the simplest aggregation and the measures of income inequality to be the most complex. The metrics discarded in this stage are assigned a redundancy measure of 1 and the remaining ones will be evaluated in the next step. 
  \item Multi-collinearity analysis: Using the aggregations obtained from the previous step, we build a linear regression model with $A_i(M)$ as the dependent variable and the remaining aggregations as predictors. We then quantify the redundancy associated with $A_i(M)$ using the adjusted coefficient of determination $R^2$. Therefore, a high value of $R^2$ would mean that aggregating $M$ using $A_i$ conveys a redundant signal as it can be well-predicted by the remaining aggregations. In this paper, we use a cutoff threshold of 0.9 to determine whether a signal is redundant or not.   
\end{enumerate}
We conduct the analysis per code metric for all aggregations. That is, for each metric, we compute the redundancy measure for all of its possible aggregations. The results we obtained are shown as box plots for three metrics: LOC in Figure \ref{boxplot_LOC}, McCabe's cyclomatic complexity in Figure \ref{boxplot_VG}, and Halstead's program level in Figure \ref{boxplot_L}. For the remaining metrics we just present the median values in Table \ref{medianRedun}.

Figure \ref{boxplot_LOC} shows that aggregating LOC using \textit{Sum} or \textit{Skew} results in unique signals as the redundancy measures for these aggregations are relatively low. For \textit{Sum}, the redundancy measures range between 0.27 and 0.58 with a median of 0.36 and for \textit{Skew} they range between 0.32 and 0.6 with a median of 0.49. On the other hand, it can be noticed that the redundancy measures associated with \textit{Kurt} and \textit{Gini} are equal to 1 in all 12 datasets. This is due to the fact that these aggregations were discarded in the correlation analysis phase. Specifically, we found that \textit{Kurt} is generally correlated with \textit{Skew} and \textit{Gini} is usually correlated with \textit{Theil}. Concerning the other aggregations, they do exhibit significantly higher redundancy measures than \textit{Sum} and \textit{Skew} but not to the extent to be considered completely redundant. In fact, each of them is deemed non-redundant (i.e. redundancy measure $< 0.9$) in at least 3 out of the 12 datasets we analyzed.

We make the same general conclusions with respect to the different aggregations of cyclomatic complexity, shown in Figure \ref{boxplot_VG}. The results, however, show that both of the income inequality measures are entirely redundant and that \textit{Kurt} exhibits some uniqueness in its signal. Concerning Halstead's program level (L), Figure \ref{boxplot_L} shows that \textit{Sum} is associated with the lowest redundancy values which range between $0.12$ and $0.23$. We also find that the measures of shape, and to a lesser extent, \textit{Avg} and \textit{SD} exhibit satisfactory uniqueness in their signals. On the other hand, \textit{Med}, \textit{IQR}, and \textit{Gini} are entirely redundant.

\begin{table}[htbp]
  \centering
  \caption{The median value of redundancy measures per code metric across the 12 datasets.}
    \begin{tabular}{rrrrrrrrrr}
    \toprule
          & Sum   & Avg   & Med   & SD    & IQR   & Skew  & Kurt  & Theil & Gini \\
    \midrule
    ev(G) & 0.45  & 0.78  & 0.53  & 1.00  & 0.62  & 1.00  & 0.32  & 1.00  & 1.00 \\
    iv(G) & 0.46  & 0.67  & 0.64  & 1.00  & 1.00  & 1.00  & 0.35  & 1.00  & 1.00 \\
    V     & 0.49  & 0.82  & 0.81  & 1.00  & 1.00  & 0.62  & 1.00  & 0.58  & 1.00 \\
    D     & 0.32  & 0.72  & 1.00  & 0.80  & 1.00  & 0.36  & 1.00  & 0.53  & 1.00 \\
    N     & 0.45  & 0.86  & 0.85  & 1.00  & 1.00  & 0.57  & 1.00  & 0.61  & 1.00 \\
    I     & 0.32  & 0.67  & 1.00  & 0.76  & 1.00  & 0.40  & 1.00  & 0.55  & 1.00 \\
    E     & 0.20  & 1.00  & 0.07  & 1.00  & 1.00  & 0.68  & 1.00  & 0.58  & 1.00 \\
    B     & 0.34  & 1.00  & 0.15  & 1.00  & 1.00  & 0.56  & 1.00  & 0.42  & 1.00 \\
    T     & 0.20  & 1.00  & 0.07  & 1.00  & 1.00  & 0.68  & 1.00  & 0.58  & 1.00 \\
    \bottomrule
    \end{tabular}%
  \label{medianRedun}%
\end{table}

Table \ref{medianRedun} shows, for each of the remaining code metrics, the median value of the redundancy measures associated with each of its aggregations across the 12 datasets. For example, the median redundancy measure of $Sum(ev(G))$ is 0.45. It can be noticed that \textit{Sum} is the only aggregation that results in a non-redundant signal for all code metrics. The corresponding redundancy measure ranges between 0.2 and 0.49. In fact, all other aggregations except \textit{Gini} result in a relatively non-redundant signal for at least one code metric. Examples include \textit{Avg(iv(G))}, \textit{Med(E)}, \textit{SD(I)}, \textit{IQR(ev(G))}, \textit{Skew(D)}, \textit{Kurt(ev(G))}, and \textit{Theil(B)}. Nevertheless, it is evident that the measures of dispersion, kurtosis and \textit{Theil} index tend to result in signals with higher redundancy levels. 

The main conclusion we draw regarding this research question is that different aggregations do convey different signals about the underlying metrics - even though some exhibit this pattern more frequently than the others. This means that code metrics could be represented at the file level by several non-redundant signals - which would be more informative than using one type of aggregation only.   

\begin{mdframed}[roundcorner=10pt]
\textit{Each of the studied aggregations, except \textit{Gini}, provides a non-redundant signal for at least one code metric. \textit{Sum} exhibits non-redundancy for all metrics.}
\end{mdframed}
\vspace{5mm}

\textbf{(RQ3) \rqthree}
\newline
\newline
The results of the two previous research questions indicate that it is possible to retain file-level aggregated metrics that are not redundant with respect to one another. Specifically, in RQ1 we found that certain aggregations do not inflate the correlation with LOC and in RQ2 we found that most studied aggregations exhibit non-redundant signals for at least one code metric. As such, we aim at investigating whether these findings could be leveraged to improve file-level defect prediction. We do so by evaluating the accuracy of defect prediction models built using each type of aggregation separately. That is, for each aggregation $A$, we build linear and logistic regression models that predict the incidence of post-release defects using the 12 code metrics aggregated to the file level using $A$. In addition, we investigate a "full" model that incorporates all the aggregations of all the code metrics at once. We denote this model by \textit{All} and use it as a means to check whether the different signals conveyed by different aggregations have an additive effect on the accuracy of defect prediction. If this is the case, then such model would outperform all models that involve one type of aggregation only. We perform predictor selection using the two approaches presented in section 3.3 and use repeated 10-fold cross validation to test our models. We measure the accuracy of the linear models using the mean-squared error of prediction and evaluate the performance of the logistic models using AUC. Table \ref{MSE} shows the average MSE that we obtained for the linear models across all the iterations of the repeated cross validation. Similarly, Table \ref{AUC} shows the average AUC for the logistic models. Rows denoted by F1 (resp. F2) correspond to the models built using the One-level (rep. Two-level) filtering approach. 

Table \ref{MSE} shows that, among the nine aggregation techniques, the linear models built using \textit{Sum} outperform those built with other aggregations for most of the datasets as they achieve lower values of MSE. We also find that aggregations pertaining to the same family yield similar results and that the measures of shape outperform the measures of dispersion and income inequality.

The results do not support the expectation that the full model would outperform those built using one type of aggregation only. When comparing the full model to the one built using \textit{Sum}, we notice a marginal improvement for the Eclipse 2.0, Eclipse 3.0, and Ant 1.7 datasets. For the remaining datasets, \textit{Sum} results in MSE values that are less than or equal to those achieved by \textit{All}. With respect to the other aggregations, the full model achieves slightly better results for all datasets except Ant 1.3, Ant 1.4, and Ant 1.5.

Another observation that could be deduced from the results is that One-level and Two-level filtering mechanisms result in similar MSE values. This means that the filtering approach does not have a significant influence on the prediction accuracy. As such, it would be obvious to favor the one that retains fewer predictor variables as it would result in a more comprehensible model that is simple to communicate. Table \ref{FilteringCount} shows the average number of variables retained by each filtering approach across the 12 datasets. The numbers adjacent to the aggregation type represent the total number of independent variables to account for. While both approaches result in major reductions, our findings indicate that Two-level filtering consistently results in fewer predictor variables and, thus, would be more suitable for model analysis.

\begin{table}[h]
\centering
\caption{Average number of variables retained by the two filtering approaches.}
\begin{tabular}{lccc}
\hline
 & One-level Filtering & Two-level Filtering\\
\hline\\[3pt]
All (108) & 18.6 & 13.8\\[3pt]
Sum (12) & 2.0 & 2.0\\[3pt]
Avg (12) & 3.6 & 2.7\\[3pt]
Med (12) & 3.7 & 3.0\\[3pt]
SD (12) & 3.7 & 2.6\\[3pt]
IQR (12) & 3.6 & 3.0\\[3pt]
Skew (12) & 3.7 & 2.7\\[3pt]
Kurt (12) & 3.2 & 2.5\\[3pt]
Theil (12) & 4.7 & 2.7\\[3pt]
Gini (12) & 4.7 & 3.0\\[3pt]
\hline
\end{tabular}
\label{FilteringCount}
\end{table}

Similarly to the linear regression results, we find that the logistic models built with \textit{Sum} generally outperform those built using other aggregations. With the exception of the Ant 1.4 dataset, \textit{Sum} achieves higher AUC values for all studied projects. \textit{Sum} also outperforms the full model for releases 1.5, 1.6, and 1.7 of Ant, as well as releases 4.1 and 4.2 of jEdit. The differences however are not significant as they range between 0.02 and 0.04. On the other hand, the full model achieves marginally better results with respect to the three releases of Eclipse in addition to releases 4.0 and 4.3 of jEdit.

In summary, Tables \ref{MSE} and \ref{AUC} show that incorporating all aggregations in the same model does not enhance the accuracy of linear or logistic defect prediction models. Specifically, we could not observe a general improvement induced by the full model (\textit{All}) with respect to the \textit{Sum} model that resulted in the best performance among the studied aggregations. This comparison was done according to the average MSE and AUC values across all the iterations of the cross validation. To further analyze the statistical significance of any potential differences between the \textit{All} and \textit{Sum} models, we use the Mann-Whitney U test. In addition to checking whether the MSE (and AUC) values exhibited by both models are statistically significant, we also quantify the extent of this difference (effect size) using Cliff's $|d|$ \citep{macbeth2011}. In this regard, \cite{romano2006} provide three thresholds to interpret the values of $|d|$: 0.147, 0.33, and 0.474. The effect would be considered negligible if it is less than 0.147, small if it is between 0.147 and 0.33, medium if it is between 0.33 and 0.474, and large otherwise. Table \ref{wilcox} shows, for each dataset, the statistical significance of the Mann-Whitney U test (p-value) in addition to Cliff's $|d|$. Concerning the results of the linear regression, we find that the difference between the two sets of MSE values is statistically significant for all datasets. However, the effect size is negligible for most of them and small with respect to Ant 1.5 and jEdit 4.3. On the other hand, the difference between the sets of AUC values exhibited by the regression models is statistically significant for half of the datasets. Among these datasets, the effect size is negligible for Eclipse 3.0 and small for the rest. These findings support our conclusion that different aggregations do not have an additive effect on the accuracy of file-level defect prediction. 

\begin{mdframed}[roundcorner=10pt]
\textit{The studied aggregations do not have an additive effect on the accuracy of defect prediction. Regression models built with all aggregations at once are not better than those built using \textit{Sum} only.}
\end{mdframed}
\vspace{5mm}

\begin{table}[htbp]
  \centering
  \caption{Average MSE achieved by the linear models across the different iterations of the cross-validation. F1 (resp. F2) corresponds to the models built using the One-level
(rep. Two-level) filtering approach.}
    \begin{tabular}{rrrrrrrrrrrr}
    \toprule
          &       & All   & Sum   & Avg   & Med   & SD    & IQR   & Skew  & Kurt  & Theil & Gini \\
    \midrule
    \multicolumn{1}{c}{\multirow{2}[1]{*}{Eclipse 2.0}} & F1    & 0.88  & 0.90  & 1.00  & 1.02  & 0.98  & 1.01  & 0.94  & 0.94  & 0.98  & 0.98 \\
    \multicolumn{1}{c}{} & F2    & 0.91  & 0.91  & 1.00  & 1.02  & 0.98  & 1.01  & 0.95  & 0.95  & 0.98  & 0.98 \\
    \hline
    \multicolumn{1}{c}{\multirow{2}[2]{*}{Eclipse 2.1}} & F1    & 0.30  & 0.30  & 0.32  & 0.32  & 0.32  & 0.32  & 0.31  & 0.31  & 0.32  & 0.32 \\
    \multicolumn{1}{c}{} & F2    & 0.30  & 0.30  & 0.32  & 0.32  & 0.32  & 0.32  & 0.31  & 0.31  & 0.32  & 0.32 \\
    \hline
    \multicolumn{1}{c}{\multirow{2}[1]{*}{Eclipse 3.0}} & F1    & 0.89  & 0.92  & 1.01  & 1.05  & 1.00  & 1.04  & 0.96  & 0.96  & 1.01  & 1.01 \\
    \multicolumn{1}{c}{} & F2    & 0.92  & 0.92  & 1.02  & 1.05  & 0.99  & 1.04  & 0.96  & 0.97  & 1.01  & 1.01 \\
    \hline
    \multicolumn{1}{c}{\multirow{2}[0]{*}{Ant 1.3}} & F1    & 0.56  & 0.47  & 0.53  & 0.51  & 0.52  & 0.51  & 0.51  & 0.53  & 0.51  & 0.52 \\
    \multicolumn{1}{c}{} & F2    & 0.54  & 0.47  & 0.52  & 0.52  & 0.51  & 0.51  & 0.50  & 0.53  & 0.53  & 0.53 \\
    \hline
    \multicolumn{1}{c}{\multirow{2}[0]{*}{Ant 1.4}} & F1    & 0.37  & 0.32  & 0.32  & 0.32  & 0.32  & 0.33  & 0.32  & 0.33  & 0.32  & 0.32 \\
    \multicolumn{1}{c}{} & F2    & 0.36  & 0.32  & 0.32  & 0.32  & 0.32  & 0.33  & 0.32  & 0.32  & 0.31  & 0.32 \\
    \hline
    \multicolumn{1}{c}{\multirow{2}[0]{*}{Ant 1.5}} & F1    & 0.13  & 0.12  & 0.14  & 0.14  & 0.14  & 0.14  & 0.13  & 0.12  & 0.13  & 0.13 \\
    \multicolumn{1}{c}{} & F2    & 0.13  & 0.12  & 0.14  & 0.14  & 0.13  & 0.14  & 0.12  & 0.12  & 0.13  & 0.13 \\
    \hline
    \multicolumn{1}{c}{\multirow{2}[0]{*}{Ant 1.6}} & F1    & 1.13  & 1.03  & 1.49  & 1.52  & 1.39  & 1.48  & 1.18  & 1.20  & 1.34  & 1.37 \\
    \multicolumn{1}{c}{} & F2    & 1.10  & 1.03  & 1.49  & 1.52  & 1.39  & 1.48  & 1.23  & 1.25  & 1.35  & 1.37 \\
    \hline
    \multicolumn{1}{c}{\multirow{2}[0]{*}{Ant 1.7}} & F1    & 1.01  & 1.04  & 1.38  & 1.42  & 1.23  & 1.42  & 1.15  & 1.19  & 1.17  & 1.24 \\
    \multicolumn{1}{c}{} & F2    & 1.05  & 1.05  & 1.38  & 1.42  & 1.25  & 1.42  & 1.15  & 1.20  & 1.19  & 1.26 \\
    \hline
    \multicolumn{1}{c}{\multirow{2}[0]{*}{jEdit 4.0}} & F1    & 4.70  & 4.61  & 6.09  & 6.08  & 6.05  & 6.15  & 5.25  & 5.54  & 6.00  & 5.81 \\
    \multicolumn{1}{c}{} & F2    & 4.78  & 4.70  & 6.09  & 6.08  & 6.00  & 6.14  & 5.08  & 5.22  & 5.95  & 5.79 \\
    \hline
    \multicolumn{1}{c}{\multirow{2}[0]{*}{jEdit 4.1}} & F1    & 3.02  & 2.86  & 3.77  & 3.85  & 3.74  & 3.87  & 3.32  & 3.70  & 3.73  & 3.61 \\
    \multicolumn{1}{c}{} & F2    & 3.10  & 2.84  & 3.77  & 3.84  & 3.75  & 3.87  & 3.31  & 3.67  & 3.73  & 3.63 \\
    \hline
    \multicolumn{1}{c}{\multirow{2}[1]{*}{jEdit 4.2}} & F1    & 1.03  & 0.96  & 1.23  & 1.25  & 1.22  & 1.25  & 1.12  & 1.06  & 1.22  & 1.19 \\
    \multicolumn{1}{c}{} & F2    & 1.08  & 0.97  & 1.23  & 1.25  & 1.22  & 1.24  & 1.10  & 1.05  & 1.21  & 1.19 \\
    \hline
    \multicolumn{1}{c}{\multirow{2}[2]{*}{jEdit 4.3}} & F1    & 0.03  & 0.03  & 0.03  & 0.03  & 0.03  & 0.03  & 0.03  & 0.03  & 0.03  & 0.03 \\
    \multicolumn{1}{c}{} & F2    & 0.03  & 0.03  & 0.03  & 0.03  & 0.03  & 0.03  & 0.03  & 0.03  & 0.03  & 0.03 \\
    \bottomrule
    \end{tabular}%
  \label{MSE}%
\end{table}

\begin{table}[htbp]
  \centering
  \caption{Average AUC achieved by the logistic models across the different iterations of the cross-validation. F1 (resp. F2) corresponds to the models built using the One-level
(rep. Two-level) filtering approach.}
    \begin{tabular}{rrrrrrrrrrrr}
    \toprule
          &       & All   & Sum   & Avg   & Med   & SD    & IQR   & Skew  & Kurt  & Theil & Gini \\
    \midrule
    \multicolumn{1}{c}{\multirow{2}[0]{*}{Eclipse 2.0}} & F1    & 0.76  & 0.75  & 0.68  & 0.64  & 0.71  & 0.66  & 0.71  & 0.66  & 0.71  & 0.71 \\
    \multicolumn{1}{c}{} & F2    & 0.75  & 0.75  & 0.69  & 0.64  & 0.71  & 0.66  & 0.71  & 0.66  & 0.70  & 0.71 \\
    \hline
    \multicolumn{1}{c}{\multirow{2}[0]{*}{Eclipse 2.1}} & F1    & 0.71  & 0.70  & 0.64  & 0.60  & 0.66  & 0.61  & 0.68  & 0.66  & 0.66  & 0.67 \\
    \multicolumn{1}{c}{} & F2    & 0.71  & 0.70  & 0.63  & 0.60  & 0.65  & 0.61  & 0.68  & 0.66  & 0.66  & 0.67 \\
    \hline
    \multicolumn{1}{c}{\multirow{2}[0]{*}{Eclipse 3.0}} & F1    & 0.75  & 0.75  & 0.65  & 0.60  & 0.67  & 0.63  & 0.72  & 0.70  & 0.66  & 0.68 \\
    \multicolumn{1}{c}{} & F2    & 0.74  & 0.75  & 0.65  & 0.60  & 0.67  & 0.63  & 0.71  & 0.69  & 0.66  & 0.68 \\
    \hline
    \multicolumn{1}{c}{\multirow{2}[0]{*}{Ant 1.3}} & F1    & 0.73  & 0.80  & 0.69  & 0.73  & 0.69  & 0.74  & 0.72  & 0.74  & 0.70  & 0.72 \\
    \multicolumn{1}{c}{} & F2    & 0.81  & 0.80  & 0.70  & 0.75  & 0.68  & 0.77  & 0.76  & 0.77  & 0.71  & 0.73 \\
    \hline
    \multicolumn{1}{c}{\multirow{2}[0]{*}{Ant 1.4}} & F1    & 0.65  & 0.64  & 0.67  & 0.66  & 0.64  & 0.63  & 0.62  & 0.64  & 0.64  & 0.65 \\
    \multicolumn{1}{c}{} & F2    & 0.64  & 0.66  & 0.67  & 0.69  & 0.67  & 0.64  & 0.67  & 0.64  & 0.69  & 0.68 \\
    \hline
    \multicolumn{1}{c}{\multirow{2}[0]{*}{Ant 1.5}} & F1    & 0.76  & 0.78  & 0.67  & 0.63  & 0.73  & 0.66  & 0.76  & 0.75  & 0.74  & 0.74 \\
    \multicolumn{1}{c}{} & F2    & 0.76  & 0.78  & 0.68  & 0.62  & 0.75  & 0.67  & 0.76  & 0.74  & 0.75  & 0.74 \\
    \hline
    \multicolumn{1}{c}{\multirow{2}[0]{*}{Ant 1.6}} & F1    & 0.80  & 0.84  & 0.67  & 0.58  & 0.73  & 0.61  & 0.80  & 0.76  & 0.74  & 0.75 \\
    \multicolumn{1}{c}{} & F2    & 0.80  & 0.84  & 0.68  & 0.58  & 0.73  & 0.62  & 0.78  & 0.75  & 0.74  & 0.74 \\
    \hline
    \multicolumn{1}{c}{\multirow{2}[0]{*}{Ant 1.7}} & F1    & 0.80  & 0.81  & 0.68  & 0.55  & 0.73  & 0.56  & 0.77  & 0.76  & 0.75  & 0.74 \\
    \multicolumn{1}{c}{} & F2    & 0.79  & 0.81  & 0.68  & 0.57  & 0.72  & 0.56  & 0.77  & 0.75  & 0.74  & 0.73 \\
    \hline
    \multicolumn{1}{c}{\multirow{2}[0]{*}{jEdit 4.0}} & F1    & 0.80  & 0.78  & 0.64  & 0.65  & 0.65  & 0.62  & 0.75  & 0.72  & 0.65  & 0.67 \\
    \multicolumn{1}{c}{} & F2    & 0.78  & 0.78  & 0.65  & 0.63  & 0.66  & 0.61  & 0.75  & 0.72  & 0.65  & 0.67 \\
    \hline
    \multicolumn{1}{c}{\multirow{2}[0]{*}{jEdit 4.1}} & F1    & 0.80  & 0.83  & 0.65  & 0.62  & 0.68  & 0.63  & 0.78  & 0.77  & 0.67  & 0.69 \\
    \multicolumn{1}{c}{} & F2    & 0.80  & 0.83  & 0.66  & 0.64  & 0.67  & 0.63  & 0.78  & 0.77  & 0.68  & 0.69 \\
    \hline
    \multicolumn{1}{c}{\multirow{2}[0]{*}{jEdit 4.2}} & F1    & 0.79  & 0.84  & 0.65  & 0.62  & 0.66  & 0.60  & 0.79  & 0.75  & 0.68  & 0.70 \\
    \multicolumn{1}{c}{} & F2    & 0.81  & 0.84  & 0.66  & 0.63  & 0.68  & 0.60  & 0.79  & 0.75  & 0.70  & 0.72 \\
    \hline
    \multicolumn{1}{c}{\multirow{2}[0]{*}{jEdit 4.3}} & F1    & 0.78  & 0.75  & 0.72  & 0.72  & 0.76  & 0.78  & 0.71  & 0.70  & 0.67  & 0.67 \\
    \multicolumn{1}{c}{} & F2    & 0.76  & 0.75  & 0.66  & 0.67  & 0.72  & 0.68  & 0.68  & 0.74  & 0.69  & 0.64 \\
    \bottomrule
    \end{tabular}%
  \label{AUC}%
\end{table}%

\begin{table}[htbp]
  \centering
  \caption{The results of Mann-Whitney U test and Cliff's delta concerning the difference in predictive accuracy between \textit{All} and \textit{Sum} models.}
    \begin{tabular}{lrrrr}
    \toprule
          & \multicolumn{2}{c}{Linear} & \multicolumn{2}{c}{Logistic} \\
    \midrule
          & p-value & Cliff's $|d|$ & p-value & Cliff's $|d|$ \\
    Eclipse 2.0 & $<0.001$ & 0.046 & $<0.001$ & 0.157 \\
    Eclipse 2.1 & $<0.001$ & 0.008 & 0.610 & 0.003 \\
    Eclipse 3.0 & $<0.001$ & 0.045 & $<0.001$ & 0.032 \\
    Ant 1.3 & $<0.001$ & 0.057 & 0.007 & 0.142 \\
    Ant 1.4 & $<0.001$ & 0.009 & 0.277 & 0.059 \\
    Ant 1.5 & $<0.001$ & 0.179 & 0.045 & 0.104 \\
    Ant 1.6 & $<0.001$ & 0.011 & $<0.001$ & 0.247 \\
    Ant 1.7 & $<0.001$ & 0.031 & $<0.001$ & 0.204 \\
    jEdit 4.0 & $<0.001$ & 0.100 & 0.833 & 0.002 \\
    jEdit 4.1 & $<0.001$ & 0.095 & $<0.001$ & 0.170 \\
    jEdit 4.2 & $<0.001$ & 0.134 & $<0.001$ & 0.169 \\
    jEdit 4.3 & $<0.001$ & 0.192 & 0.766 & 0.019 \\
    \bottomrule
    \end{tabular}%
  \label{wilcox}%
\end{table}%

\section{Threats to Validity}
\label{sec:threats}

\subsection{Internal Validity}
Our analysis considers a rather small subset of static static code metrics. Therefore, the conclusions that we draw might not generalize to other types of software metrics or even to other code metrics. Concerning the aggregation techniques that we employed, they are not comprehensive although they cover a wide range of commonly used measures. We also applied aggregation at the file level only. Higher level aggregations might have different implications.

\subsection{External Validity}
We used three open source projects to analyze the impact of aggregation on defect prediction. This choice was constrained by the availability of file-level defect information. Although the studied systems are mature and commonly used in literature, our findings might not apply to other systems.

\section{Conclusion}
\label{sec:conclusions}

While defect prediction is usually conducted at the file level, many code metrics are defined at the method level. To overcome this problem, researchers often aggregate such metrics using summation to build file-level defect prediction models. Previous research has shown that summation-based aggregation increases correlation between code metrics, which is likely to render many of them redundant at the file level. In this paper, we investigated nine different aggregation techniques relative to their impact on defect prediction. In addition to summation, we explored measures of central tendency, dispersion, and income inequality. The set of code metrics that we used included lines of code, McCabe, and Halstead metrics.

Using 12 releases of three open source projects, our results indicate that different aggregation techniques do convey different statistical information but their collective impact on defect prediction does not outperform simple summation. Specifically, we found that certain aggregations such as \textit{Median} do not inflate file-level correlation with LOC and that most of the studied aggregations provide a non-redundant signal for at least one code metric. However, when building logistic and linear defect prediction models, we found that models that incorporate more types of aggregations do not have a higher predictive power than those built using summation only.

As future work, we intend to investigate whether our findings hold for higher levels of aggregation, e.g. package, plugin, etc. We also plan to study the impact of aggregation on other types of software metrics such as process metrics \citep{rahman2013} and ownership metrics \citep{bird2011}. As opposed to code metrics which measure different aspects of the code structure, these metrics relate to the history of changes associated with a software artifact. They are usually defined at the file level but could be defined at a higher level if need be.

\begin{acknowledgements}
I would like to thank Dr. Ahmed E. Hassan and Mr. Shane McIntosh for helping make this work possible.
\end{acknowledgements}

\bibliographystyle{spbasic}
\bibliography{references}

\end{document}